%% file: sb1608a.tex
\documentclass[english, oldversion]{aa}
\usepackage{mathptmx}
\usepackage[T1]{fontenc}
\usepackage[latin9]{inputenc}
\usepackage{float}
\usepackage{graphicx}
\usepackage{amssymb}
\usepackage[authoryear]{natbib}

\makeatletter

\providecommand{\tabularnewline}{\\}

\def\bron{4U~1608-522}
\input{aas_macros.tex}

\newcommand{\ee}[1]{\ensuremath{\cdot 10^{#1}}}
\newcommand{\unitspace}{\ensuremath{\,}}
\newcommand{\usp}{\unitspace}
\newcommand{\numberspace}{\ensuremath{\;}}
\newcommand{\nsp}{\numberspace}
\newcommand{\unitstyle}[1]{\ensuremath{\mathrm{#1}}}
\newcommand{\power}[2]{\ensuremath{{#1}^{#2}}}
\newcommand{\centi}{\unitstyle{c}}
\newcommand{\Mega}{\unitstyle{M}}
\newcommand{\meter}{\unitstyle{m}}
\newcommand{\cm}{\centi\meter}
\newcommand{\gram}{\unitstyle{g}}
\newcommand{\second}{\unitstyle{s}}
\newcommand{\Kelvin}{\unitstyle{K}}
\newcommand{\K}{\Kelvin}  %degrees Kelvin
\newcommand{\grampersquarecm}{\gram\usp\power{\cm}{-2}} %column depth
\newcommand{\columnunit}{\grampersquarecm}
\newcommand{\erg}{\unitstyle{erg}} %ergs
\newcommand{\ergspersecond}{\erg\unitspace\power{\second}{-1}}
\newcommand{\amu}{\unitstyle{u}} %atomic mass unit
\newcommand{\eV}{\unitstyle{eV}}        %eV
\newcommand{\MeV}{\Mega\eV} %MeV
\newcommand{\yr}{\unitstyle{yr}}        %year
\newcommand{\days}{\unitstyle{d}}
\newcommand{\grampersecond}{\gram\usp\power{\second}{-1}}

\newcommand{\nuclei}[2]{\ensuremath{\mathrm{^{#1}#2}}}
\newcommand{\carbon}[1][12]{\nuclei{#1}{C}}

\usepackage{babel}
\makeatother

\begin{document}
\authorrunning{L. Keek et al.} \titlerunning{First superburst from a classical low-mass X-ray binary transient}

\title{First superburst from a \\
 classical low-mass X-ray binary transient}

\author{L.~Keek\inst{1,2}, J.~J.~M.~in~'t~Zand\inst{1,2}, E.~Kuulkers\inst{3},
A.~Cumming\inst{4}, E.~F.~Brown\inst{5}, M.~Suzuki\inst{6}}

\institute{SRON Netherlands Institute for Space Research, Sorbonnelaan 2, NL -
3584 CA Utrecht, the Netherlands \and Astronomical Institute, Utrecht
University, P.O. Box 80000, NL - 3508 TA Utrecht, the Netherlands
\and ISOC, ESA/ESAC, Urb. Villafranca del Castillo, P.O. Box 50727,
28080 Madrid, Spain \and  Physics Department, McGill University,
3600 rue University, Montreal, QC, H3A 2T8, Canada\and  Department
of Physics \& Astronomy, National Superconducting Cyclotron Laboratory,
and the Joint Institute for Nuclear Astrophysics, Michigan State University,
East Lansing, MI 48824 \and  Institute of Space and Astronautical Science, JAXA,
1-1, Sengen 2chome, Tsukuba-city, Ibaraki 305-8505, Japan}

\offprints{L. Keek, email {\tt l.keek@sron.nl}.}

\date{Received ; Accepted}

\abstract{We report the analysis of the first superburst from a transiently
accreting neutron star system with the All-Sky Monitor (ASM) on the Rossi
X-ray Timing Explorer. The superburst occurred 55~days after the
onset of an accretion outburst in \object{\bron}. During that time interval,
the accretion rate was at least 7\% of the Eddington limit. The peak
flux of the superburst is 22 to 45\% of the Eddington limit, and
its radiation energy output is between $4\cdot10^{41}$ and $9\cdot10^{41}$~erg
for a distance of 3.2~kpc. Fits of cooling models to the superburst
light curve indicate an ignition column depth between $1.5\cdot10^{12}$ and $4.1\cdot10^{12}$~g~cm$^{-2}$.
Extrapolating the accretion history observed by the ASM, we derive that this
column was accreted over a period of 26 to 72 years.
The superburst characteristics are consistent with
those seen in other superbursting low-mass X-ray binaries. However, the transient
nature of the hosting binary presents significant challenges for superburst theory,
requiring additional ingredients for the models.
The carbon that fuels the superburst is thought to be produced mostly during
the accretion outbursts and destroyed in the frequent type-I X-ray bursts.
Mixing and sedimentation of the elements in the neutron star envelope may
significantly influence the balance between the creation and destruction of carbon. Furthermore,
predictions for the temperature of the neutron star crust fail to reach the values 
required for the ignition of carbon at the inferred column depth.
}

\maketitle

\section{Introduction\label{intro}}

\bron\ is a bright low-mass X-ray binary (LMXB) with a neutron-star
(NS) primary (\citealt{bel76,tab76}). QX~Nor was identified by \citet{GrindlayLiller1978}
as the optical counterpart. The accretion onto the neutron star is
transient (\citealt{loc94} and references therein) and the outburst
behavior complex. Three main flux states can be resolved in \bron\ (\citealt{wac02}):
an outburst state, a low-intensity state and a true quiescent state.
In contrast to many other LMXB transients, its true quiescent state
has a relatively low duty cycle of only about 50\%. Wachter et al.
speculate that the mass transfer rate might be fluctuating around
the critical threshold separating stable accretion systems from true
transient systems, possibly due to varying stellar spots on the donor
star or to a varying vertical accretion disk structure which may change
the irradiation and, thus, the mass transfer rate through the disk.
\citet{wac02} also identified a modulation in the optical light curve
with a period of 12.9 hr which they suspect to be near the orbital
period. Thus, the donor is probably hydrogen-rich.

\bron\ exhibits so-called type-I X-ray bursts (\citealt{bel76,tab76}).
These are understood to result from thermonuclear shell flashes on
the surfaces of NSs of layers of accreted hydrogen and helium (\citealt{woo76,mar76};
for reviews, see \citealt{lew93} and \citealt{str06}). The flash
layers heat up during a fraction of a second and cool down in an exponential-like
manner lasting 10\,s to a few minutes. As a result, a spectrum may
be observed which is typically well fit by a black body with a temperature
decreasing from a peak of typically $\mathrm{k}T=2$ to 3 keV and
an emission region similar in size to what is expected for a NS (\citealt{swank1977}).
Some nuclear energy generation may persist throughout the cooling
phase, prolonging the bursts somewhat. A number of bursts exhibit
photospheric radius expansion (PRE) where the burst flux is thought
to reach the Eddington limit. From the observation of PRE bursts from
\bron, \citet{gal07} derived a distance to this source of $3.2\pm0.3$~kpc.
During 5 PRE bursts, burst oscillations were detected at 619~Hz, which implies
the fastest known spin frequency for accreting neutron stars (\citealt{gal07}),
with the possible exception of XTE~J1739-285 (\citealt{Kaaret2007}).

Since a few years a different kind of type-I X-ray burst is detected
in about 10\% of all bursters: so-called `superbursts' (\citealt{cor00,str02}).
These are about $10^{3}$ times more energetic and last for hours
to a day. They are thought to result from flashes of carbon-rich layers
(\citealt{str02,2001CummingBildsten}). These layers are presumed
to be much further down in the neutron star than those where ordinary
X-ray bursts occur, which would explain the longer duration. It has
been observed that superbursts influence the normal bursting behavior.
Each time when the start of a superburst has been observed with sufficient
statistics a precursor burst was seen (see e.g. \citealt{str02}).
Furthermore, after the superburst the normal bursting behavior is
quenched for approximately one month (see e.g. \citealt{Kuulkers2002ks1731}).

Superbursts have so far only been seen in systems in which the NS
has been continuously accreting for at least 10~yr (\citealt{Kuulkers:2003wt,2004intZand}),
although not in every such system (\citealt{kee06}). Here we present
a superburst from \bron, which represents the first case for a `classical'
transient. A superburst has also been observed from the transient
source KS~1731-260 (\citealt{Kuulkers2002ks1731}). However, unlike
KS~1731-260, \bron~exhibits states of high flux with a duration
that is short with respect to the expected superburst recurrence time.
This may have interesting consequences for the theory of superbursts,
because the different states of the transient allow us to study different
levels of carbon production and burning in one source. Perhaps most importantly, whereas the crust in KS~1731-260 is expected to have been heated significantly out of thermal equilibrium with the core during its long outburst, the crust of \bron\ is not expected to be significantly heated during its short outbursts. This presents a significant challenge to carbon ignition models for superbursts, which require a hot crust to achieve ignition at the depths inferred from observations of superbursts. Despite these differences, the superburst from \bron\ has similar properties to superbursts from other sources.

The paper is organized as follows. First we describe the observations and spectral
calibration in Sect. \ref{sec:Observations-and-spectral}. In Sect.
\ref{bursts} we report on an analysis of the X-ray burst behavior
of \bron\ as a function of its apparent mass accretion rate. Next,
in Sect. \ref{sb}, we present the analysis of the detected superburst.
Finally, we discuss the implications of the results for superburst
theory in Sect. \ref{discussion}. Initial reports of this superburst
appeared in \citet{rem05} and \citet{kuu05}.

\section{Observations and spectral calibration\label{sec:Observations-and-spectral}}

\subsection{Observations}

To study both the long-term accretion and bursting behavior of \bron,
as well as the superburst, we employ observations performed with multiple
instruments on-board four X-ray observatories.

The All-Sky Monitor (ASM) on the Rossi X-ray Timing Explorer (RXTE)
(\citealt{Levine1996}) consists of three Scanning Shadow Cameras
(SSC), each containing a position-sensitive proportional counter.
The cameras are mounted on a rotating drive. Data are accumulated
in so-called dwells of 90 seconds. After each dwell the rotation drive
changes the orientation of the SSCs. Spectral information is available
from three channels with corresponding energy ranges of roughly 1.5--3,
3--5 and 5--12 keV. After the launch of RXTE in December 1996 until November 2006, the
ASM observed \bron~for a total of 2.8~Ms. On May 5th 2005 the ASM
observed the superburst from \bron.

RXTE also carries the Proportional Counter Array (PCA). The PCA consists
of five proportional counter units (PCUs) with a total geometric area
of 8000 cm$^{2}$ and has a bandpass of 1 to 60 keV (\citealt{Jahoda2006}).
\bron~was observed for 1.6~Ms in total, mostly when it was out
of quiescence (e.g., \citealt{vst03,Gierlinkski2002}).

The BeppoSAX observatory (\citealt{1997Boella}) was launched in April
1996 and carried two Wide Field Camera's (WFCs) as well as four Narrow
Field Instruments (NFI). The WFCs (\citealt{Jager1997}) are coded
mask aperture cameras with a band pass of 2--28~keV. During the six
year lifespan of BeppoSAX a campaign of semi-yearly observations of
the Galactic Center were carried out, which resulted in an exposure
time of 3.8~Ms for \bron.

The BeppoSAX NFI performed two pointed observations of \bron. The
first observation was performed during an outburst on February 28
1998. In this paper we analyze the
broad-band spectrum obtained with the following NFI in
the indicated energy bands: the Low Energy Concentrator Spectrometer
(LECS; \citealt{1997Parmar}; 0.12--4~keV), the Medium Energy Concentrator
Spectrometer (MECS; \citealt{1997BoellaMECS}; 1.8--10~keV) and the
Phoswich Detection System (PDS; \citealt{1997Frontera}; 15--220~keV).
The exposure time depends on the instrument and is 30\,ks for the MECS.
The second observation took place when \bron~was in quiescence.
Since we are interested in the production of carbon from accreted
matter, we forgo the analysis of the data from this observation.

On May 5th 2005, 33 minutes before the start of the ASM observation
of the superburst, the High Energy Transient Explorer 2 (HETE-2; \citealt{Ricker2003})
observed a flare from \bron~with two instruments: the Wide field
X-ray Monitor (WXM; \citealt{Shirasaki2003}) with a 2 to 25~keV band
pass and the French Gamma Telescope (FREGATE; \citealt{Atteia2003})
which has a band pass of 6 to 400~keV.

The International Gamma-Ray Astrophysics Laboratory (INTEGRAL; 
\citealt{Winkler2003}) was launched in October 2002. The
Imager on Board the Integral Satellite (IBIS; \citealt{Ubertini2003}) 
is a coded aperture camera containing two detectors. We use results 
obtained with the INTEGRAL Soft Gamma-Ray Imager (ISGRI; \citealt{Lebrun2003})
detector, which has an energy range from 15 to about 500~keV. IBIS/ISGRI 
has observed \bron\ up to September 2005 for a total of 6.3~Ms.

In this paper we analyze spectra using version 11.2.0 of the XSPEC
software package (\citealt{Arnaud1996}).

\subsection{Spectral calibration of the ASM\label{sub:Spectral-calibration-of}}

The spectral response of the ASM detectors is not well defined. A
pre-flight effective area array is available, while a redistribution
matrix is not (see discussion in  \citealt{Kuulkers2002}). We construct
such a matrix by modeling, for an infalling photon with a certain
energy, the energy distribution over the three channels by a Gaussian
with a full width at half maximum of 20\% of the centroid energy,
which is a typical width for the type of proportional counter used
in the ASM (e.g. \citealt{Fraser1989}). The response matrix thus
obtained is merely a first order estimate, but suffices for rough
calculations.

We employ the Crab source to investigate the accuracy of our response
matrix. The average count rate over all ASM observations from this
source in the full 1.5--12 keV band-pass is $75.4\,\mathrm{c\, s^{-1}}$.
The X-ray spectrum can be described by an absorbed power law. \citet{Kirsch2005}
performed simultaneous model fits to Crab spectra obtained with 22
X-ray instruments. Using the results of these fits in the 2--10 keV
range, our response matrix%
\footnote{\citet{Kuulkers2002} finds the count rate is over-predicted. This
is due to a different choice of values for the spectral model parameters.%
} predicts an ASM 1.5--12 keV count rate of $70.5\,\mathrm{c\, s^{-1}}$.
Therefore, when performing spectral analyses using this matrix, the
normalization of the models needs to be corrected by a factor $1.07$
to account for this discrepancy. Note that in principle this factor
can be different for each of the three SSCs and can vary with time.
However, we find that in a time interval of 100 days centered at the
start of the superburst the difference in Crab count rate between
any two SSCs never exceeds the $3\sigma$ level.

Apart from the accuracy in predicting the count rate of the Crab,
we also investigate how well our response matrix can reproduce the
typical model parameters that are found by fitting an absorbed power
law to the three-channel spectral data. We extracted a spectrum from
all the ASM data on the Crab available at the time of writing. We
fix the hydrogen absorption column density at the value of $N_{\mathrm{H}}=0.45\cdot10^{22}\,\mathrm{cm^{-2}}$,
as found by \citet{Kirsch2005} (in the 0.1--1000 keV energy range).
Following \citeauthor{Kirsch2005}, we use the abundances found by
\citet{Wilms2000} and cross sections from \citet{Verner1996}. Leaving
free the power law index and the normalization, we do not find an
acceptable agreement with the data. Only if we add in quadrature 10\%
of the flux to the uncertainty of each data point are we able to obtain
an acceptable fit with $\chi_{\mathrm{red}}^{2}\simeq1$. Taking into
account the correction factor derived previously, we find the best
fit with a photon index of $\Gamma=2.01\pm0.12$ and a normalization
of $N_{\mathrm{powerlaw}}=9\pm2$ photons $\mathrm{keV^{-1}cm^{-2}s^{-1}}$,
which is consistent with the results from \citet{Kirsch2005}. The
uncertainty in $N_{\mathrm{powerlaw}}$ is large, because by definition
$N_{\mathrm{powerlaw}}$ is the photon flux at 1 keV, which is outside
of the ASM energy range. The XSPEC power law model `pegpwrlw' uses
a user-defined energy range for the normalization. Employing the ASM
bandpass gives an uncertainty in the normalization of 6\%.

\section{History of accretion and X-ray burst activity\label{bursts}}

\subsection{Long-term light curve\label{sub:Long-term-light-curve}}

\begin{figure}[t]
\includegraphics[width=1\columnwidth]{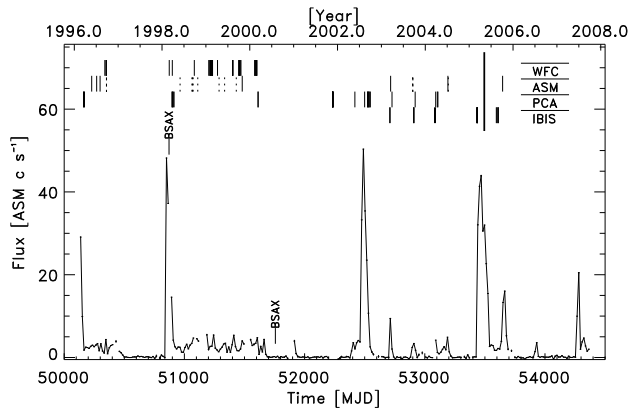}

\caption{1.5--12 keV RXTE-ASM light curve of \bron\ at a 2-week time resolution.
Data points with errors in excess of 0.5~c~s$^{-1}$ were excluded
from this plot. The connecting line is broken if data points are more
than 2 weeks apart. The vertical lines indicate 37 bursts detected
with the WFC, 19 with the ASM (dashed lines indicate tentative bursts),
31 with the PCA (\citealt{gal07}) and 19 with the IBIS/ISGRI (\citealt{Chelovekov2007}). The long
vertical line indicates the time of the superburst. `BSAX' indicates
the times of two BeppoSAX observations. \label{figasm}}

\end{figure}

Figure~\ref{figasm} shows the 1.5-12 keV ASM light curve of \bron.
Clearly visible are the three states identified by \citeauthor{wac02}
(2002; see Sect. \ref{intro}). Four major outbursts, where the peak photon
count rates was in excess of 20\,c~s$^{-1}$ for more than one month, are visible in the 11-year
time span of the observations (see Sect. \ref{sub:Accretion-rate}
for a discussion on the selection criterion). The first outburst was
ongoing at the start of the ASM observations and lasted until 60 days
afterward, with fluxes in excess of $20$\,c~s$^{-1}$. The following
outbursts lasted with fluxes above $20$\,c~s$^{-1}$ for 44~d (MJD
50848-50892), 48~d (MJD 52475-52523), and 80~d (MJD 53438-53518),
respectively. There are also five minor outbursts, after the two latter
major outbursts. The low-intensity states are visible in the first
half of the mission. From comparing the first half and the latter
half of the light curve, it appears that either a low-intensity state
emerges after a major outburst, or a series of minor outbursts. This
characteristic is only now apparent, after 11 years of observation
(Wachter et al. 2002 only considered the first 5 years of the data
set). We determined the average flux over the complete ASM data to
be $2.466\pm0.006\,\mathrm{c\,s^{-1}}$. This is about 2\% of the
flux reached during the brightest bursts seen in the ASM.

\subsection{Accretion rate\label{sub:Accretion-rate}}

When matter is accreted onto the neutron star surface, some of the
accretion energy and/or some of the material may leave in a jet. If
these losses are significant, the persistent flux is not a good tracer
of the mass accretion rate $\dot{M}$. For black-hole X-ray binaries,
\citet{Fender2005} argue that jets are not present during the high
soft state. It is during the high soft state of \bron~that we are
particularly interested in the precise value of the accretion rate.
Therefore, assuming that the same holds true for neutron star binaries,
in this paper we will assume there are no losses through a jet and
that the persistent flux is a good measure for $\dot{M}$. Expressed
as a fraction of the Eddington-limited mass accretion rate $\dot{M}_{\mathrm{Edd}}$,
which is $\dot{M}_{\mathrm{Edd}}=2\cdot10^{-8}M_{\odot}\mathrm{yr^{-1}}$
for a canonical hydrogen-accreting neutron star with a mass of $1.4{\, M}_{\odot}$, 
it can simply be derived as $\dot{M}/\dot{M}_{\mathrm{Edd}}=F/F_{\mathrm{Edd}}$,
with $F$ the bolometric flux and $F_{\mathrm{Edd}}$ the Eddington-limited
flux. We will determine the bolometric flux below. The Eddington-limited flux is exhibited
during photospheric radius expansion (PRE) bursts. RXTE PCA observed
12 such bursts (\citealt{gal07}). The mean unabsorbed bolometric peak flux of
these bursts is $F_{\mathrm{Edd}}=(1.32\pm0.14)10^{-7}\,\mathrm{erg\, s^{-1}\, cm^{-2}}$.

Van Straaten et al. \citeyearpar{vst03} make an extensive study of
the spectral changes of \bron\ during the various states and the
associated timing properties. They conclude that \bron\ is an Atoll
source that exhibits the `banana' state during outbursts and moves
in the color-color diagram to the `extreme island' state when the
flux goes to levels of the low-intensity state. The distinguishing
parameter is the `hard color' (i.e., the flux between 9.7 and 16.0
keV divided by the flux between 6.0 and 9.7 keV) which is about 1.15
in the extreme island state and 0.55 in the banana state. This implies
that one-band measurements, such as the ASM count rates, are not an
accurate measure of the bolometric flux, let alone the accretion rate.
In order to derive the bolometric flux during the low and high accretion
states, we analyze the broad-band outburst-observation with BeppoSAX
as well as a large number of observations with the RXTE PCA.

Two broad-band observations were carried out during the ASM coverage
with BeppoSAX, one of which during outburst (for the timing of these
observations, see Fig.~\ref{figasm}). To determine the flux during
this observation we analyze spectra obtained with the LECS, MECS and
PDS. We subtract background spectra taken from blank field observations
at the same detector positions for the LECS and MECS spectra, while
for the PDS off-source pointings are used as background. The spectra
are rebinned to obtain at least 15 photons in each bin while sampling
the instrument resolution with at most 3 channels at all energies,
to ensure the applicability of the $\chi^{2}$ statistic. An error
of 1\% is added in quadrature to the statistical error per bin to
account for systematic uncertainties. This is common practice in
BeppoSAX analyses (\citealt{Fiore1999}). We fit the data from all
three instruments simultaneously with the generic LMXB model (see
e.g. \citealt{2001Sidoli}), consisting of a multi-temperature disk
black body (\citealt{1984Mitsuda,1986Makishima}; `diskbb' in XSPEC)
in combination with a comptonized spectrum (\citealt{1994Titarchuk,1995Hua,1995Titarchuk};
`comptt' in XSPEC), both absorbed by cold interstellar matter following
the model by \citet{Balucinska1992} (`phabs' in XSPEC). During the
spectral analysis we allow the normalization between the three instruments
to vary and find that the best fit values are within acceptable limits:
the normalization between the LECS and MECS is $0.897\pm0.003$, while
0.7 to 1.0 is acceptable; between the PDS and MECS the normalization
is $1.21\pm0.04$, while 1.1 to 1.3 is acceptable (\citealt{Fiore1999}).
The results of the fit as well as the derived flux are provided in
\begin{table}
\caption{\label{cap:Flux-determination-with}Parameters and $1\sigma$
uncertainties of fitting the outburst spectrum of \bron~from BeppoSAX NFI
observations on 28 Feb 1998 with an absorbed disk black
body$^a$ and comptonized$^b$ model.}

\begin{center}
\begin{tabular}{ll}
\hline 
$N_{\mathrm{H}}$ & $\left(8.91\pm0.05\right)10^{21}\,\mathrm{cm^{-2}}$\tabularnewline
 & \tabularnewline
$\mathrm{k}T_{\mathrm{BB}}$ & $2.38\pm0.07$ keV\tabularnewline
$N_{\mathrm{BB}}$ & $18.3\pm0.9$\tabularnewline
 & \tabularnewline
$\mathrm{k}T_{0}$ & $0.478\pm0.005$ keV\tabularnewline
$\mathrm{k}T_{e}$ & $3.6\pm0.2$ keV\tabularnewline
$\tau$ & $3.7\pm0.2$\tabularnewline
$N_{\mathrm{Comptt}}$ & $0.40\pm0.02$\tabularnewline
 & \tabularnewline
$\chi_{\mathrm{red}}^{2}/\mathrm{dof}$ & $1.40/160$\tabularnewline
Flux (0.01--300 keV)$^c$ & $\left(1.75\pm0.10\right)10^{-8}\,\mathrm{erg\, s^{-1}\, cm^{-2}}$\tabularnewline
ASM count rate$^d$ & $39.4\pm0.9\,\mathrm{c\, s^{-1}}$\tabularnewline
\hline
\end{tabular}

\par\end{center}

$^a$ Disk black body model parameters: $\mathrm{k}T_{\mathrm{BB}}$ 
is the temperature at the inner disk radius $R_{\mathrm{in}}$ and
$N_{\mathrm{BB}}\equiv(R_{\mathrm{in}}^{2}/d_{10}^{2})\cos\theta$
the normalization, with $R_{in}$ in units of km, $d_{10}$
the source distance in units of 10 kpc and $\theta$ the inclination.

$^b$ Comptonized model parameters: $\mathrm{k}T_{0}$, $\mathrm{k}T_{e}$ are resp.
the seed photon and plasma temperature, $\tau$ the plasma optical
depth for a disk geometry and $N_{\mathrm{Comptt}}$ the normalization.

$^c$ Unabsorbed 0.01--300 keV flux, which we take to be bolometric.

$^d$ Average ASM count rate during the observation.

\end{table}
Table~\ref{cap:Flux-determination-with}. The flux is determined
by extrapolating the model from the observed range to 0.01--300 keV.
Extrapolating the model further while correcting for the interstellar
absorption does not lead to a significant increase in flux. Under
the assumption that no other features in the spectrum other than those
we modeled have a substantial flux contribution, we refer to this
as the bolometric flux. From the average ASM flux during this observation,
we find that 1 ASM c~s$^{-1}$ is equivalent to $(4.4\pm0.2)\cdot10^{-10}$~erg~s$^{-1}$cm$^{-2}$.

\begin{figure}
\begin{centering}
\includegraphics[width=0.7\columnwidth]{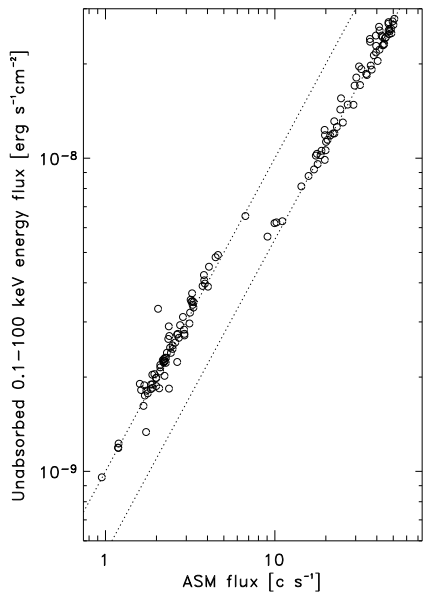}
\par\end{centering}

\caption{\label{fig:Flux-as-a}Flux as a function of the ASM count rate. The
distinction between the two spectral states is clearly visible. The
two lines are not fits, but act as guides to distinguish the two states.}

\end{figure}

To further study the bolometric correction that may be applied to
the ASM data as a function of the persistent flux, we employ 160 observations
carried out with the RXTE PCA between March 1996 and February 2006.
The PCA spectra are extracted from the standard products, with exposure
times between 96 s and 13 ks and a total exposure of 310 ks. They
are modeled between 3 and 40 keV (the calibrated bandpass) with a
generic model consisting of a black body and a cutoff power law, both
absorbed following the model by \citet{1983Morrison} with $N_{{\rm H}}=8\cdot10^{21}$~cm$^{-2}$.
The resulting fits are acceptable in 144 cases with $\chi_{{\rm red}}^{2}<2$;
in the remaining cases $\chi_{{\rm red}}^{2}$ rises to up to 6. Despite
the sometimes bad fit, we employ all observations since we are interested
only in obtaining constraints for the bolometric correction and not
in physical implications of the spectral model. The fit outcome is
tested leaving $N_{{\rm H}}$ free: no evidence is found for fluctuations
in $N_{{\rm H}}$ above $5\cdot10^{22}$~cm$^{-2}$. We determine
the photon flux between 1.5 and 12 keV before correcting for absorption
and normalize it to observed ASM fluxes through calibration with Crab
fluxes and by cross correlating the PCA measurements for \bron~with
contemporaneous ASM fluxes. We calculate the 0.1 to 100 keV energy
flux after a correction for absorption. Obviously, this implies a
fairly large extrapolation from the 3 to 40 keV PCA bandpass and we
check how much flux is found outside that bandpass. It varies between
10\% at the brightest state to 60\% in the faintest state. This indicates
what the maximum error is of the extrapolation. The results are plotted
in Fig.~\ref{fig:Flux-as-a}. It clearly shows the two states identified
by \citet{vst03}. The boundary between both states is at 5 to 10
ASM c~s$^{-1}$. At the bright state, when the spectrum is relatively
soft, 1 ASM c~s$^{-1}$ is equivalent to $5.5\cdot10^{-10}$~erg~s$^{-1}$cm$^{-2}$
and at the hard faint state to 1.0$\cdot10^{-9}$~erg~s$^{-1}$cm$^{-2}$.
For the bright state we can calibrate this result using the BeppoSAX
broadband observation of the 1998 outburst from which we derived above
the equivalence of 1 ASM c~s$^{-1}$ to $(4.4\pm0.2)\cdot10^{-10}$~erg~s$^{-1}$cm$^{-2}$.
We use this number as conversion factor for the high flux state. %

\begin{table*}

\caption{\label{tab:Outbursts-properties-with}Outbursts properties. Between parentheses we give the  1$\sigma$
uncertainties in the least significant digit(s). }

\begin{center}
\begin{tabular}{lllllll}
\hline 
Outburst & Start time$^{\mathrm{a}}$ & Duration$^{\mathrm{a}}$ & Fluence & Energy$^{\mathrm{b}}$ & Peak Flux $\mathrm{^{c}}$ & Peak $\dot{M}/\dot{M}_{\mathrm{Edd}}$ $\mathrm{^{d}}$\tabularnewline
 & (MJD) & (days) & ($10^{-2}\,\mathrm{erg\, cm^{-2}}$) & ($10^{44}$erg) & ($10^{-8}\,\mathrm{erg\, s^{-1}\, cm^{-2}}$) &\tabularnewline
\hline
1 & 50848.0 & 44.0 & $6.3(3)$ & $0.77(15)$ & $3.0(2)$ & $0.23(3)$\tabularnewline
2 & 52475.0 & 48.0 & $6.9(3)$ & $0.85(16)$ & $2.6(2)$ & $0.20(3)$\tabularnewline
3 & 53437.5 & 79.5 & $10.4(5)$ & $1.3(2)$ & $2.24(19)$ & $0.17(2)$\tabularnewline
3 before SB & 53437.5 & 57.6 & $8.2(4)$ & $1.01(19)$ & $2.24(19)$ & $0.17(2)$\tabularnewline
\hline
\end{tabular}
\par\end{center}

$^{\mathrm{a}}$ Start time and duration of time interval when the
outburst flux exceeds 20 ASM c s$^{-1}$.

$^{\mathrm{b}}$ Energy calculated from fluence using distance $d=3.2\pm0.3$
kpc.

$\mathrm{^{c}}$ Unabsorbed bolometric peak flux determined from the 1-day ASM light curve
($10^{-8}\,\mathrm{erg\, s^{-1}\, cm^{-2}}$).

$\mathrm{^{d}}$ Peak mass accretion rate in units of the Eddington limit assuming a 1.4 M$_{\odot}$ neutron star.\end{table*}

Applying the bolometric correction factors to the ASM flux, we find
that the average persistent flux was $(2.313\pm0.013)10^{-9}\,\mathrm{erg\, cm^{-2}\, s^{-1}}$,
while the peak flux reached during the three large outbursts was $2.2\cdot10^{-8}$
to $3.0\cdot10^{-8}$ erg cm$^{-2}$ s$^{-1}$. The mass accretion
rate is on average $\dot{M}/\dot{M}_{\mathrm{Edd}}=(1.8\pm0.2)10^{-2}$
and $0.17$ to $0.23$ in the peaks of the outbursts.

Using the conversion factor of ASM counts to bolometric flux for count
rates above 10 c s$^{-1}$ we determine the fluence of the three major
outbursts observed by the ASM for the duration that the count rate
exceeded 20 c s$^{-1}$. For the last outburst we also determined
the fluence up to the moment that the superburst occurred. 20 c s$^{-1}$
corresponds to an accretion rate of $\dot{M}/\dot{M}_{\mathrm{Edd}}=(6.7\pm0.8)10^{-2}$
and is chosen as a lower limit since it is close to the value of 0.1
that is required for the production of carbon (\citealt{2001CummingBildsten}),
while selecting the periods of high flux in one interval for each
outburst. The bolometric fluences as well as the peak fluxes during
the outbursts are presented in Table \ref{tab:Outbursts-properties-with}.

\subsection{Long-term bursting behavior\label{sub:Long-term-bursting-behavior}}

Indicated in Fig.~\ref{figasm} are the bursts detected from \bron\ with
the WFC on BeppoSAX until the mission ended in April 2002, the IBIS/ISGRI on
INTEGRAL until September 2005 (\citealt{Chelovekov2007}), the PCA
(\citealt{gal07}) and the ASM on RXTE. One burst was detected by
both the WFC and PCA. The ASM bursts are identified through \emph{1)}
searching individual 90-sec average dwell data of \bron~for isolated
high points that are at least $3\sigma$ above a 100-day running average,
resulting in roughly 200 burst candidates; \emph{2)} studying the
raw 1-sec resolution light curves of the candidate dwells, to eliminate
candidates that have no clear flare feature with a time scale smaller
than the 90-s dwell time; \emph{3)} searching for flares with clear
fast-rise exponential-decay profiles, resulting in the identification
of 7 certain bursts. The remaining 12 candidates are classified as
tentative bursts from \bron.

\begin{figure}
\includegraphics[width=1\columnwidth]{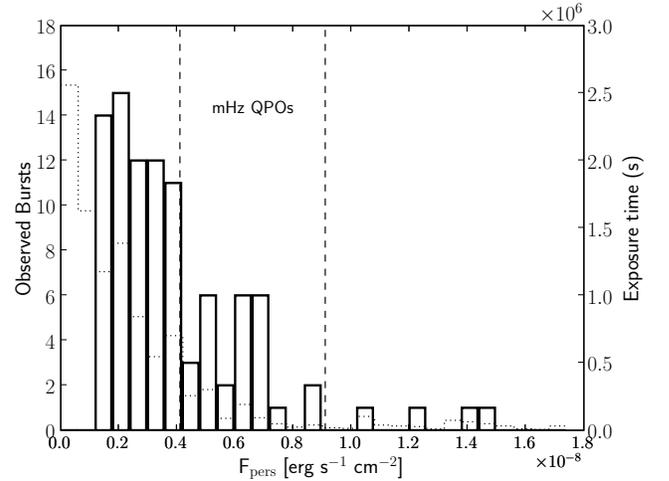}

\caption{\label{fig:Number-of-observed}Number of observed bursts ({\em solid line}) as a function
of persistent flux $F_{\mathrm{pers}}$ . For the latter we use the
average ASM count rate in a one-day interval around the moment the
burst occurred converted to bolometric flux. This includes all bursts
observed with BeppoSAX WFC, RXTE PCA, INTEGRAL IBIS/ISGRI as well as 
the clear bursts
identified in the RXTE ASM light curve. The {\em dashed lines} indicate
the region of $F_{\mathrm{pers}}$ where \citet{Revnivtsev2001} observed
mHz QPOs (see Sect. \ref{sub:Bursts}). The {\em dotted line} represents the 
total PCA, WFC, IBIS/ISGRI and ASM exposure time as a function of the ASM count
rate. For PCA, WFC and IBIS/ISGRI observations we use the average ASM count rate
during an observation. Together with the burst histogram, one observes that
the burst rate does not vary significantly as a function of accretion rate in
the range where type-I bursts occur.}

\end{figure}

For studying the bursting behavior as a function of the persistent
flux, Fig. \ref{fig:Number-of-observed} %
shows a histogram of the ASM-measured persistent flux when a burst
is seen for all bursts observed with the WFC, PCA and IBIS/ISGRI as well as the
certain bursts observed with the ASM. There is the suggestion that
bursts are rare during major outbursts while much more frequent during
the extended low-intensity states during MJD 50150--50500 and 50897--51800.
Quantitatively, however, the data are inconclusive in this respect.
The PCA data set is somewhat biased, because those observations were
targeted at \bron\ mainly when it was in outburst. The other data
sets are serendipitous and, therefore, present an objective view of
the bursting behavior. The WFC exposure time during major outbursts
is 21.3 times smaller than during low-intensity states, while the
ratio in the number of bursts is 1 to 36. 2 out of 19 bursts observed by
IBIS/ISGRI take place during a major outburst, while the source is observed
in outburst during 12.7\% of the exposure time. The ASM data are also 
inconclusive:
2 out of 19 bursts occurred at 1-day average fluxes of 10\,c\,s$^{-1}$
while the fraction of the total time spent in that flux regime is
7.4\%. The WFC, PCA, IBIS/ISGRI and ASM light curves exhibit bursts with an average
recurrence time of, respectively,  $1.19$, $0.60$, $3.79$ and $1.71$~days.

The bursting behavior is characterized by $\alpha$: the ratio of
persistent fluence between bursts and the burst fluence. Using the
PCA observations we calculate the average value of $\alpha$ both
for the outbursts, when the ASM count rate was in excess of 20\,c\,s$^{-1}$,
and for all observations when the ASM count rate was below this level, 
but not during quiescence.
\citet{gal07} determined the fluence of 31 of the type-I X-ray bursts
observed with the PCA. We find the persistent fluence between bursts
by multiplying the average ASM flux during the PCA observations by
the burst recurrence time. For the latter we take the ratio of the
total PCA exposure time and the number of observed bursts during either
the outbursts or the low-flux state. During the outbursts the PCA
observed only one burst. Therefore, from the PCA data we only have
a lower limit to the burst recurrence time in the outbursts, from
which we find the lower limit $\alpha\geq745\pm9$. The PCA observations
when the ASM flux was less than 20\,c\,s$^{-1}$ yield an average value
of $\alpha=51.5\pm0.2$.

\section{Superburst analysis\label{sb}}

\subsection{Light curve}

\begin{figure}
\includegraphics[width=1\columnwidth]{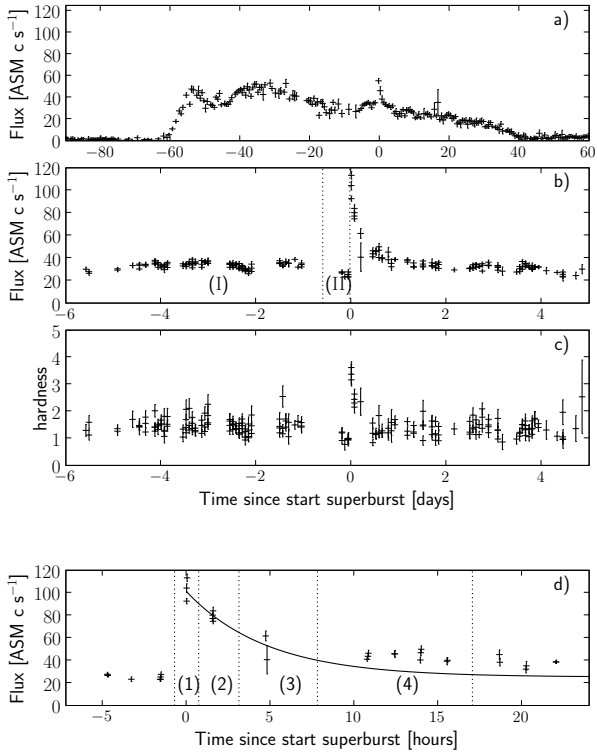}

\caption{\label{fig:Superburst-lightcurve}Superburst light curves where each
data point represents a 90\,s dwell with its $1\sigma$ uncertainty.
\emph{a)} 1.5--12\emph{~}keV ASM count rate at a 0.5 day time resolution.
Shown is the outburst in which the superburst takes place. \emph{b)}
Zoomed-in part of \emph{a}. Indicated are intervals I and II of the
persistent flux prior to the burst. \emph{c)} hardness ratio for each
dwell, defined as the ratio of the counts in the 5--12~keV and the
1.5--3~keV energy bands. One data point with an error in excess of
4 is excluded from the plot. \emph{d)} Zoomed-in part of top figure
on the superburst. Burst intervals 1--4 are indicated as well as the
exponential decay fitted to the first 5 hours of the burst with an
e-folding decay time of 4.8 hr (see Table \ref{tab:Superburst-properties.}).}

\end{figure}

\begin{figure}
\includegraphics[width=1\columnwidth]{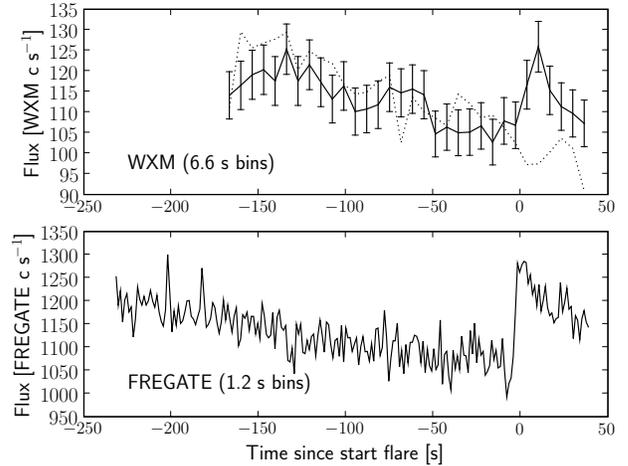}

\caption{\label{fig:HETE-2-light-curve}HETE-2 light curve of likely onset
superburst. {\em Top}: WXM. The data points show the count rate with $1\sigma$ uncertainty
observed by detector XB, while the dotted line indicates the count
rate observed by detector XA. We employ data from the full 2--25~keV
band pass at a 6.6\,s time resolution. The data is from the end of
an observation, when the count rate decreases as the Earth covers
an increasing part of the field of view. The flare is only visible
in the XB light curve. From its position on the sky, \bron\ was in
the field of view of XB and not of XA. {\em Bottom}: FREGATE light curve
from the 6--40~keV band pass at 1.2\,s time resolution.}

\end{figure}

In Fig. \ref{fig:Superburst-lightcurve} %
we show the ASM light curve of \bron\ for the outburst in which the
superburst takes place. The onset of the superburst is not observed
by the ASM, as it falls in a 1.5 hour data gap. Thirty-three minutes
before the start of the superburst observation by the ASM, the light
curves of the WXM and FREGATE onboard HETE-2 exhibit a fast rise of
3.6\,s followed by a slow decay (Fig. \ref{fig:HETE-2-light-curve}).
Since the event took place at the end of the observation, only one
minute of the flare is recorded before the instruments are switched
off. Nevertheless, the light curve suggests that the decay is much longer than
for a normal type-I burst. The WXM detector image of the flare is 
consistent with the position
of \bron~if the spacecraft's attitude changed by $0.6^{\circ}$
since the middle of the orbit. A variation of the attitude of this
size is not uncommon at the end of an orbit. The FREGATE has no positional
information, but \bron\ was in the field of view at that time. From both
instruments spectral information is available in four energy bands. Due to
the low number of observed photons per energy band, this information is of
limited use for the WXM. For the FREGATE, the flare is observed in the 6 to
40~keV and 6 to 80~keV bands and not in higher energy bands, which is
consistent with a type-I X-ray burst (such as a superburst).
Comparing this flare
to a type-I X-ray burst from \bron\ which was previously observed
with the WXM, the flare's peak height is approximately 80\% of that
of the burst. Therefore, the peak flux of the flare is at most 80\%
of the Eddington limit. As the time of occurrence, the source position
and the peak flux are consistent with the superburst from \bron,
we regard this as a likely observation of the start of this superburst.

No precursor burst can be discerned. However, the presence of such
a burst cannot be ruled out. If a potentially present precursor is
similar to the type-I bursts observed during outbursts, it would have a
short e-folding decay time of a few seconds and a peak flux of around 60\%
of the Eddington limit. Before the superburst onset the statistical quality of 
the FREGATE data is such that we can exclude at a $3\sigma$ confidence 
level a precursor burst with a decay time of several seconds and a net
peak flux of $\sim 10$\% of the Eddington limit or higher. However, the
precursors observed from other hydrogen-accreting superbursters all took
place at the superburst onset (\citealt{Kuulkers2002ks1731,Strohmayer2002,2003intZand}). 
If the precursor and superburst light curves
are superimposed, the presence of a precursor cannot be excluded from the data. Also, during the 
precursor burst the peak temperature might be lower than for the
superburst, which would result in a lower observed flux in the FREGATE band. The WXM
is sensitive down to lower energies, but the relative uncertainty in each data point
is larger due to a smaller effective area, which may prevent us from detecting a precursor.

After the superburst 
took place, the first normal type-I X-ray burst was observed with IBIS/ISGRI after
$99.8$ days. Since there are frequent data gaps, whose durations
are long with respect to the duration of an X-ray burst, 
$99.8$ days is an upper limit to the burst quenching time.

In the 4.6 hours preceding the superburst (interval II, see Fig. \ref{fig:Superburst-lightcurve}b)
the persistent flux is at a level of $25.3\,\mathrm{c\, s^{-1}}$,
while in the 7 days prior to that (interval I) it was $32.9\,\mathrm{c\, s^{-1}}$.
The hardness ratio (Fig. \ref{fig:Superburst-lightcurve}c) indicates that
the spectrum is harder in interval I.
Approximately one day after the superburst peak, the count rate returns
to the level of interval I. Only after 4.5 days does the count rate return
to the level of interval II. This suggests that the superburst starts during
a slight dip in the persistent flux. This makes it difficult to accurately
separate the persistent emission from the burst emission during the
decay. Therefore, we report results using both interval I and II.

The decay of superbursts is typically well-fit by an exponential.
We fit the decay of the superburst with an exponential in all three
channels separately as well as in the combined 1.5--12 keV energy
band (see Table \ref{tab:Superburst-properties.} and Fig. \ref{fig:Superburst-lightcurve}d).
In this fit we restrict ourselves to the first 5 hours of the decay,
to avoid the tail where an assessment of the persistent emission is
difficult.%
\begin{table}
\caption{\label{tab:Superburst-properties.}Superburst properties using persistent
interval I and II with $1\sigma$ uncertainties for the last digits
between parentheses.}

\begin{tabular}{lllll}
\hline 
\multicolumn{5}{c}{Exponential decay times$^{a}$}\tabularnewline
Persistent: & I &  & II & \tabularnewline
\hline
Band (keV) & $\tau_{\mathrm{exp}}$ (hr) & $\chi_{\mathrm{red}}^{2}$/d.o.f. & $\tau_{\mathrm{exp}}$ (hr) & $\chi_{\mathrm{red}}^{2}$/d.o.f.\tabularnewline
\hline
1.5--12 & $4.1_{-0.5}^{+0.6}$ & 6.2/7 & $4.8_{-0.4}^{+0.5}$ & 6.5/7\tabularnewline
1.5--3  & $12.2_{-6.9}^{+-}$ & 1.6/6 & $13.2_{-6.2}^{+-}$ & 1.6/6\tabularnewline
3--5 & $7.8_{-1.7}^{+3.3}$ & 1.4/7 & $7.8_{-1.4}^{+2.3}$ & 1.5/7\tabularnewline
5--12 & $3.3_{-0.4}^{+0.5}$ & 2.7/7 & $3.9_{-0.3}^{+0.4}$ & 3.2/7\tabularnewline
Bolometric & $3.0_{-0.3}^{+0.4}$ & 4.3/7 & $3.6_{-0.3}^{+0.3}$ & 5.7/7\tabularnewline
\hline 
\multicolumn{5}{c}{Persistent emission prior to flare ($N_{\mathrm{H}}$ coupled)$^{b}$}\tabularnewline
Interval$^{c}$ & I & II &  & \tabularnewline
\hline
$N_{\mathrm{H}}$ & \multicolumn{2}{c}{$6.1(4)\cdot10^{21}\,\mathrm{cm^{-2}}$} &  & \tabularnewline
$\Gamma$ & $1.75(3)$ & $2.06(7)$ &  & \tabularnewline
$N_{\mathrm{powerlaw}}$ & $2.8(2)$ & $3.3(4)$ & \multicolumn{2}{l}{photons $\mathrm{keV^{-1}cm^{-2}s^{-1}}$}\tabularnewline
$\chi^{2}$/d.o.f. & \multicolumn{2}{c}{$1.7/1$} &  & \tabularnewline
Flux$^{d}$ & $1.3(3)$ & $1.0(3)$ & \multicolumn{2}{l}{$10^{-8}\,\mathrm{erg\, s^{-1}\, cm^{-2}}$}\tabularnewline
\hline 
\multicolumn{5}{c}{Superburst spectral analysis ($N_{\mathrm{bb}}$ coupled)$^{e}$;
persistent interval I}\tabularnewline
Interval$^{c}$ & 1 & 2 & 3 & \tabularnewline
\hline
$\mathrm{k}T$ (keV) & $2.14(14)$ & $1.90(12)$ & $1.60(8)$ & \tabularnewline
$N_{\mathrm{bb}}$ & \multicolumn{4}{c}{$1.4(3)\cdot10^{2}$}\tabularnewline
$\chi_{\mathrm{red}}^{2}$/d.o.f. & \multicolumn{4}{c}{$3.6/5$}\tabularnewline
Flux$^{f}$ & $3.18(6)$ & $1.96(4)$ & $1.00(3)$ & \tabularnewline
\hline 
\multicolumn{5}{c}{Superburst spectral analysis ($N_{\mathrm{bb}}$ coupled)$^{e}$;
persistent interval II}\tabularnewline
Interval$^{c}$ & 1 & 2 & 3 & 4\tabularnewline
\hline
$\mathrm{k}T$ (keV) & $2.13(12)$ & $1.89(9)$ & $1.67(8)$ & $1.38(7)$\tabularnewline
$N_{\mathrm{bb}}$ & \multicolumn{4}{c}{$1.7(3)\cdot10^{2}$}\tabularnewline
$\chi_{\mathrm{red}}^{2}$/d.o.f. & \multicolumn{4}{c}{$3.5/7$}\tabularnewline
Flux$^{f}$ & $3.81(2)$ & $2.37(2)$ & $1.438(11)$ & $0.68(6)$\tabularnewline
\hline 
\multicolumn{5}{c}{Peak flux and burst fluence; persistent interval I}\tabularnewline
Burst start: & ASM & HETE-2 & Maximum & \tabularnewline
\hline
Flux$_{\mathrm{peak}}$$^{f}$ & $2.90(14)$ & $3.5(3)$ & $4.8(6)$ & \tabularnewline
Fluence $^{g}$ & $3.1(4)$ & $3.8(6)$ & $5.2(1.0)$ & \tabularnewline
$y_{12}$$^{h}$ & $1.5$ & $1.7$ & $2.1$ & \tabularnewline
$E_{17}$$^{i}$ & $1.4$ & $1.6$ & $1.9$ & \tabularnewline
\hline
\multicolumn{5}{c}{Peak flux and burst fluence; persistent interval II}\tabularnewline
Burst start: & ASM & HETE-2 & Maximum & \tabularnewline
\hline
Flux$_{\mathrm{peak}}$$^{f}$ & $3.88(11)$ & $4.5(2)$ & $6.0(5)$ & \tabularnewline
Fluence $^{g}$ & $5.0(4)$ & $5.9(5)$ & $7.6(1.1)$ & \tabularnewline
$y_{12}$$^{h}$ & $3.0$ & $3.3$ & $4.1$ & \tabularnewline
$E_{17}$$^{i}$ & $1.5$ & $1.6$ & $1.9$ & \tabularnewline
\hline
\end{tabular}

$^{a}$In the 1.5--3 keV band we find the decay to be consistent with
a horizontal line, due to the large uncertainty of the data points
in the tail of our fit interval.

$^{b}$Absorbed power law model ($N_{\mathrm{powerlaw}}E^{-\Gamma}$)
parameters: $N_{\mathrm{H}}$ the interstellar column density, $\Gamma$
the photon index and $N_{\mathrm{powerlaw}}$ the normalization defined
as photons $\mathrm{keV^{-1}cm^{-2}s^{-1}}$ at 1 keV.

$^{c}$See Fig. \ref{fig:Superburst-lightcurve}b.

$^{d}$ Unabsorbed flux in 1.5--12 keV energy range ($10^{-8}\,\mathrm{erg\, s^{-1}\, cm^{-2}}$).

$^{e}$ Black body model parameters: temperature $\mathrm{k}T$ and
normalization $N_{\mathrm{bb}}\equiv R^{2}/d^{2}$, with R the source
radius in km and d the source distance in 10 kpc.

$^{f}$ Unabsorbed bolometric flux ($10^{-8}\,\mathrm{erg\, s^{-1}\, cm^{-2}}$).

$^{g}$ Unabsorbed bolometric fluence ($10^{-4}\,\mathrm{erg\, cm^{-2}}$).

$^{h}$ Column depth ($10^{12}\,\mathrm{g\, cm^{-2}}$).

$^{i}$ Energy release per unit mass ($10^{17}\,\mathrm{erg\, g^{-1}}$).
\end{table}

Not only is the count rate in interval II lower than in I. Also, the
hardness ratio indicates that the spectrum is softer in interval II
(see ).

\subsection{Spectral analysis, energetics and layer thickness\label{sub:Spectral-analysis}}

We fit the persistent
emission ASM spectra from both interval I and II simultaneously with an absorbed
power law, coupling the $N_{\mathrm{H}}$ value for both spectra.
The best fit values as well as the 1.5--12 keV flux in
each interval are provided in Table \ref{tab:Superburst-properties.}.

We divide the superburst in four time intervals, as indicated in Fig.
\ref{fig:Superburst-lightcurve}d, from the part of the burst where
most data points lie at $>3\sigma$ above the persistent flux from
interval II (when using interval I, we restrict ourselves to the first
three burst intervals). These intervals contain in total 19 measurements.
Note that some of these start at the same time, but are from different
SSCs. A systematic error of 3\%, derived from the scatter in the Crab
intensity (\citealt{Levine1996}), is included in the uncertainties
of the data points. \citet{Levine1996} note that in many cases this
is an underestimate of the systematic uncertainties. For each interval
we extract a spectrum. We obtain the net-superburst spectrum by subtracting
the persistent spectrum based on either interval I or II from the
observed spectrum (see \citealt{Kuulkers2002gx17+2} for a discussion
on decoupling persistent and burst emission for normal type-I bursts).
This net spectrum is fit with an absorbed black body.

During the spectral analysis we fix the hydrogen absorption column
density. From BeppoSAX broad-band observations during outburst we
find a hydrogen absorption column density of $N_{\mathrm{H}}=0.89\cdot10^{22}\,\mathrm{cm^{-2}}$
(see Sect. \ref{sub:Accretion-rate}). However, here we use the value
of $N_{\mathrm{H}}=0.61\cdot10^{22}\,\mathrm{cm^{-2}}$ which we obtained
from the persistent emission in interval II, since we perform the
superburst spectral analysis using similar ASM data and the same response
matrix.

In the spectral analysis, our best fit has a reduced $\chi_{\mathrm{red}}^{2}$
value of 3.5. We increased the uncertainties in the data points such
that we obtained a fit with $\chi^{2}=1$. Using this fit we determined
the uncertainties in the fitted parameters using $\Delta\chi^{2}=1$.
Increasing the uncertainties to satisfy the {}``goodness of fit''
condition $\chi_{\mathrm{red}}^{2}$ was also required when we analyzed
the Crab spectrum in Sect. \ref{sub:Spectral-calibration-of}. In
this procedure the best fit values of the model parameters deviate
from the previous best fit values by at most $1.37\sigma$. We fit
the four intervals simultaneously, keeping the value of the black
body normalization the same for all spectra. Since the normalization
parameter depends on the emission region size of the black body, this
implies that the size is assumed not to change during the superburst
decay. This is a valid approach, as we will see that the peak flux
of the superburst is below the Eddington flux at which photospheric
expansion is expected. Fits where we did not use this coupling did
not indicate a significant change in radius from one interval to the
next. We provide the best fit values of the model parameters in Table
\ref{tab:Superburst-properties.}. Assuming isotropic radiation and
using the distance of $3.2\pm0.3$~kpc as well as the best fit value
for the black body normalization, we find an emission region size
equivalent to a sphere with a radius of $R=4.2\pm0.5$~km, which
is consistent with the black body radii both \citet{Murakami1980} and \citet{gal07} found
for normal type I X-ray bursts.

To investigate the influence of the choice of $N_{\mathrm{H}}$ on
the fit results, we repeated the spectral analysis using $N_{\mathrm{H}}=8.91\cdot10^{21}\,\mathrm{cm^{-2}}$
(see Table \ref{cap:Flux-determination-with}). The difference in
best fit values of the model parameters to those in the previous analysis
never exceeds $1\sigma$. The derived flux differs by at most 3\%.

Using the average count rate and the flux we obtained for each burst
interval, we convert the count rate of each data point to flux. We
fit an exponential function to the first 5 hours of the resulting
curve and determine the peak flux at the start of this interval. We
find a peak flux of $22\pm3\%$ ($29\pm3$\%) of the Eddington flux
using the persistent level from interval I (II). Integrating this
exponential gives us the burst fluence, which at a distance of $3.2\pm0.3$~kpc
corresponds to a burst energy of $(3.8\pm0.9)10^{41}\,\mathrm{erg}$
($(6.1\pm1.2)10^{41}\,\mathrm{erg}$). By extrapolating the exponential
to the start of the flare observed by HETE-2, a peak flux of $27\pm4\%$
($34\pm4\%$) of the Eddington flux and a burst fluence of $(4.7\pm0.8)10^{41}\,\mathrm{erg}$
($(7.2\pm0.9)10^{41}\,\mathrm{erg}$) are obtained. Extrapolating
to the start of the ASM data gap preceding the observed part of the
burst, we find an upper limit to the peak flux and burst fluence:
respectively $36\pm6$\% ($45\pm4$\%) of the Eddington flux and $(6\pm2)10^{41}\,\mathrm{erg}$
($(9\pm2)10^{41}\,\mathrm{erg}$) (see Table \ref{tab:Superburst-properties.}).

\begin{figure}
\includegraphics[bb=81bp 206bp 545bp 655bp,width=1\columnwidth]{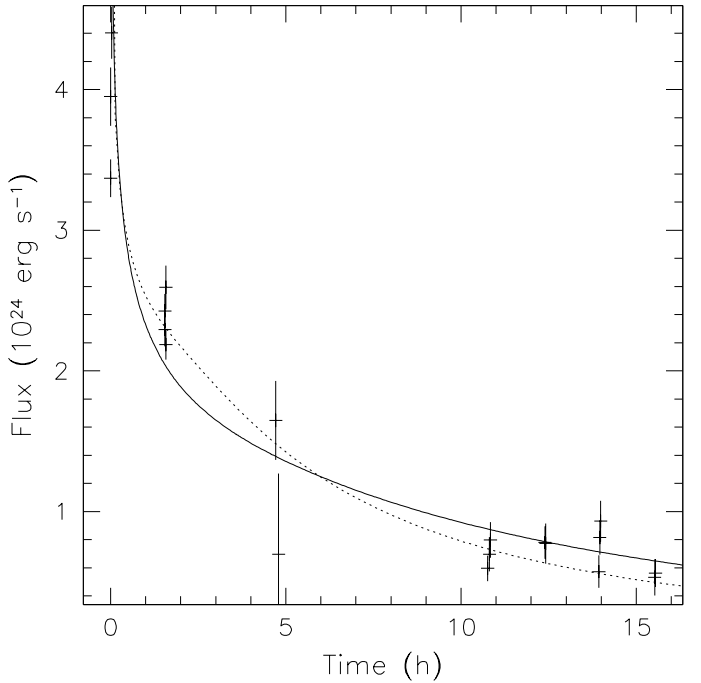}

\caption{\label{fig:Fit-of-cooling}Fit of cooling models to superburst decay.
The data points are ASM count rates converted to flux for each ASM
measurement (see Sect. \ref{sub:Spectral-analysis}) with $1\sigma$
uncertainties. The persistent level from interval II is used (see Fig.
\ref{fig:Superburst-lightcurve}b). The solid and dotted line represent
fits with, respectively, the numerical and analytic cooling models
from \citet{CummingMacbeth2004}.}

\end{figure}

\begin{table}
\caption{\label{tab:Characteristics-of-the}Characteristics of the \bron~superburst
compared to 13 other superbursts (from \citealt{Kuulkers:2003wt}
and \citealt{2004intZand}). $\tau_{\mathrm{exp}}$ from fit to count
rates converted to unabsorbed bolometric flux.}

\begin{tabular}{lll}
\hline 
 & \bron & Other superbursts\tabularnewline
Duration (hr) & 5--22 & 2--15\tabularnewline
Precursor & No & Always when data available\tabularnewline
$\tau_{\mathrm{exp}}$(hr)$^{a}$ & $\sim4.5$ & 0.7--6\tabularnewline
$\mathrm{k}T_{\mathrm{peak}}$(keV) & $2.1\pm0.1$ & 1.8--3\tabularnewline
$L_{\mathrm{peak}}(10^{38}\,\mathrm{erg\, s^{-1}})$ & 0.4--0.7 & 0.4--3.4\tabularnewline
$L_{\mathrm{pers}}(L_{\mathrm{Edd}})$ & 0.11--0.15 & 0.1--0.25$^{b}$\tabularnewline
$\langle L_{\mathrm{pers}}\rangle (L_{\mathrm{Edd}})$ & 0.03 & 0.11--0.43$^{c}$\tabularnewline
$E_{\mathrm{b}}(10^{42}\,\mathrm{erg})$ & 0.4--0.9 & 0.3--1.4\tabularnewline
$\tau\equiv E_{\mathrm{b}}/L_{\mathrm{peak}}(\mathrm{hr})$ & 1.0--1.3 & 1.1--6.9\tabularnewline
$\gamma\equiv L_{\mathrm{pers}}/L_{\mathrm{peak}}$ & 0.2--0.5 & 0.1--0.7\tabularnewline
$t_{\mathrm{no\, bursts}}(\mathrm{d})$ & <99.8 & >7.5\tabularnewline
Donor & H/He & He and H/He\tabularnewline
\hline
\end{tabular}

$^{a}$ Exponential decay time in 1.5--12 keV band for \bron~and
in similar energy bands for other superbursters.

$^{b}$ Persistent luminosity at the time of the superburst, excluding the candidate
 ultracompact source 4U~0614+091 and the Z-source GX~17+2 which have a luminosity close
to, respectively, 1\% (\citealt{kuu05,intZand2007}) and 100\% (\citealt{2004intZand}) of the Eddington luminosity.

$^{c}$ Average persistent luminosity from all ASM observations, again excluding 4U~0614+091 and GX~17+2.
\end{table}
\citet{CummingMacbeth2004} present multizone numerical models for
the cooling of the neutron star surface layers during the decay of
a superburst as well as analytic fits to these models. The models
depend on the ignition column depth $y\equiv y_{12}10^{12}\mathrm{g\, cm^{-2}}$
and the energy release per gram $E\equiv E_{17}10^{17}\mathrm{erg\, g^{-1}}$.
\citet{Cumming2006} fit these models to the light curves of six superbursts.
We perform the same fit to the light curve of the superburst from
\bron~using the persistent level from both interval I and II (Fig.
\ref{fig:Fit-of-cooling}%
). We find, respectively, $y_{12}=1.5$, $E_{17}=1.4$ and $y_{12}=3.0$,
$E_{17}=1.5$. To estimate the values of $y_{12}$ and $E_{17}$ in
case the superburst started at either the start of the flare observed
with HETE-2 or at the start of the data gap in the ASM light curve,
we use the scaling laws found by \citet{Cumming2006}. Thus we obtain
the ranges $y_{12}=1.5-2.1$, $E_{17}=1.4-1.9$ and $y_{12}=3.0-4.1$,
$E_{17}=1.5-1.9$, using respectively the persistent level from interval
I and II (see Table \ref{tab:Superburst-properties.}).

\section{Discussion\label{discussion}}

The first detection of a superburst from a classical LMXB transient
poses significant challenges to superburst theory and underlines 
shortcomings thereof that were already realized earlier. This shows
the need for additional ingredients in the models. While the superburst
characteristics of \bron\ are similar to those from other superbursters
(see Table  \ref{tab:Characteristics-of-the}), the accretion rate on a 
time scale of ten years is a factor of 3 smaller than observed for the slowest accretor
among all other superbursting sources.

In the following subsections we will see that there are two challenges: the destruction
of carbon by the ever-present hydrogen/helium flashes is apparently not as
important as theory predicts and the predicted crust temperatures fall short
by a factor of 2 of the value required for the ignition of unstable carbon
burning. Additionally, we make a few remarks on the mHz QPO observed in \bron.

\subsection{Recurrence time}

From fits of cooling models to the superburst light curve we find
that the superbursting layer had at ignition a column depth of $y_{12}=1.5-4.1$
and an energy release per unit mass of $E_{17}=1.5-1.9$, which implies
a total energy release of $(2.6-9.8)10^{42}\,\mathrm{erg}$. From
the X-ray flux we derived an energy release of $(0.4-0.9)10^{42}\,\mathrm{erg}$,
which is only $5-45$\% of the total inferred from the cooling models
(after correction for the gravitational redshift with a typical value
of $1+z=1.3$). The rest of the energy is either released as neutrinos
or is conducted to greater depths and radiated away on a longer
timescale. %
\begin{figure}
\includegraphics[width=1\columnwidth]{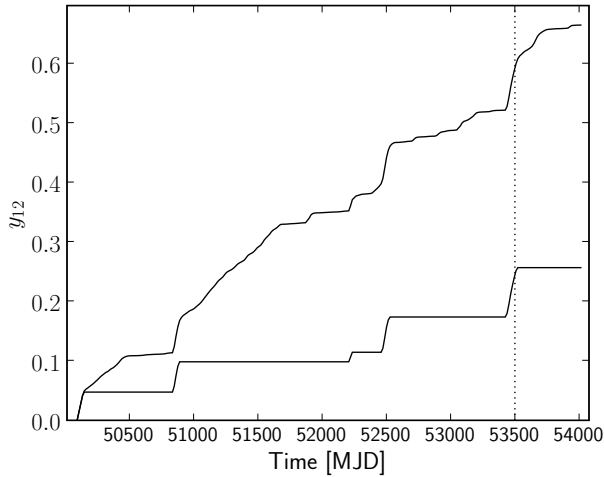}

\caption{\label{fig:Accreted-column-depth}Accreted column depth $y_{12}\equiv y/(10^{12}\mathrm{g\, cm^{-2}})$
over time. Assuming a neutron star of $1.4\,\mathrm{M}_{\odot}$ mass
and $10$ km radius at a distance of $3.2$ kpc, we show the cumulative
column depth that is accreted since the start of RXTE ASM observations
as inferred from the X-ray flux at 14 day time resolution (c.f. Fig.
\ref{figasm}). The lower curve takes into account only matter that
is accreted when \bron~was in the high state when the ASM count
rate exceeded 20\,c\,s$^{-1}$, while the upper curve includes the
low state as well. The dotted line indicates the moment the superburst
was observed.}

\end{figure}
 Fig. \ref{fig:Accreted-column-depth} shows the column depth that
is observed to be accumulated since the start of the ASM observations.
At the moment of the superburst a column of $y_{12}=0.55$ ($y_{12}=0.27$
taking into account only the stable burning in the high state) had
been accreted over a period of 9.3 years. Note that this is derived
under the assumption that the X-ray flux is a good measure for the
mass accretion rate, which may not be the case (see discussion in
Sect.~\ref{sub:Accretion-rate}). In fact, we assume that stable
burning occurs when the ASM flux is in excess of 20 c\ s$^{-1}$,
which introduces an uncertainty of several tens of percents in the
column depth. With an average accretion rate of $\dot{M}=(1.8\pm0.2)10^{-2}\,\dot{M}_{\mathrm{Edd}}$
the required column depth is accreted over a period of 26 to 72 years.
If carbon is only produced from matter accreted during stable burning
in the high state, the recurrence time is twice as long. 
In both cases it is much longer than the 57.6 day duration
of the 2005 outburst before the superburst occurred. This implies
that the carbon fuel has been produced during multiple outbursts.

\subsection{Producing and preserving carbon}

\citet{Schatz2003} found from nucleosynthesis calculations that carbon is produced most
efficiently through the stable nuclear burning of hydrogen and helium. Unstable burning during type-I X-ray bursts results in significantly less carbon ($\lesssim 1$\% by mass; \citealt{Woosley2004}), but produces heavier ashes. \citet{2001CummingBildsten} proposed that these heavy ashes are important because they set the opacity and therefore temperature of the superbursting layer, although \citet{Brown2000} later showed that this effect is much smaller than initially estimated once the temperature profile of the entire crust was taken into account. Therefore, it is interesting to ask whether both burning regimes are necessary for producing the fuel for superbursts. Certainly, all known superbursters exhibit both type-I X-ray bursts and a high $\alpha$ value, which are indications for both unstable and stable burning, respectively. 

The stable burning of helium takes place at accretion rates $\dot{M}$ in excess of 10\% of Eddington.
Therefore, we expect that in \bron~carbon is created during the
outbursts. This carbon could be destroyed by type-I X-ray bursts, which are observed
during all flux states. For carbon to survive the bursts, it has to
reside in a layer which is depleted of hydrogen. \citet{Woosley2004}
find this to be the case at the bottom of a bursting layer. The different composition of the ashes of unstable hydrogen/helium burning as opposed to stable hydrogen/helium burning (\citealt{Woosley2004,Schatz2003}) will likely mix rapidly. If a period of stable burning was followed
by unstable burning, the heavier ashes of the latter would rapidly
undergo a Rayleigh-Taylor instability and mix with the lighter ashes
below.

An important question is whether the carbon is able to burn stably during the low state. \citet{2001CummingBildsten} argued that carbon would burn stably at low accretion rates, and suggested that this may explain why superburst sources have accretion rates of $\approx0.1$ Eddington and up. This question is crucial to understand for \bron, since the accretion rate is low during quiescence and the low persistent state. Further modelling of the production and destruction of carbon during the cycle of transient outbursts is needed.

A complication here is that at the low accretion rates that occur during quiescence and in the
low persistent state in \bron, the accretion may be slow enough that
the carbon and heavy elements in the ashes of normal type-I bursts
gravitationally separate, with the carbon floating upwards relative
to the heavy elements. To estimate the relative drift velocity of
a carbon nucleus in the heavy element ocean, we follow the work of
\citet{BildstenHall2001}, who estimated the rate of sedimentation
of $^{22}$Ne in the interiors of CO white dwarfs. They start with
the self-diffusion coefficient for a one component plasma, $D\approx3\omega_{p}a^{2}\Gamma^{-4/3}$
(\citealt{Hansen1975}), where $\omega_{p}$ is the ion plasma frequency,
$a$ is the inter-ion separation, and $\Gamma$ is the standard parameter
that measures the strength of the Coulomb interactions between ions.
A single carbon nucleus in a background of $^{56}$Fe feels an upward
force of $12m_{p}g/13$, giving a drift velocity $v=54m_{p}g/169e\Gamma^{1/3}(4\pi\rho)^{1/2}$
(compare \citealt{BildstenHall2001} eq.~{[}3]). Writing the time
to cross a pressure scale height at this drift velocity as $t_{s}=H/v=P/\rho gv$,
and assuming that the pressure is provided by degenerate relativistic
electrons (a good approximation at these depths), we find \[
t_{s}=7.2\ {\rm years}\ y_{12}^{13/24}\ \left(\frac{T_{8}}{6}\right)^{-1/3}\left(\frac{g_{14}}{2.45}\right)^{-31/24},\]
 where we assume that the heavy element is $^{56}$Fe (the scaling
is $t_{s}\propto Z^{37/18}/A$). This timescale is equal to the accretion
timescale $t_{{\rm accr}}$ for $\dot{m}\approx0.05\ \dot{m}_{{\rm Edd}}$.
The sedimentation timescale $t_{s}$ is comparable to the timescale
on which we infer the carbon fuel is accumulated, so that the physics
of the relative separation of the carbon and heavy elements should
be considered further.

\subsection{The crust temperature and carbon ignition\label{s.crust-temperature}}

Although the accretion rate of \bron~reaches $\dot{M}\la0.23\dot{M}_{\mathrm{Edd}}$
during the peak of the outburst, the high accretion rate only lasts
a short ($\la100\nsp\days$) time, so that the crust does not reach
a thermal steady-state. The temperature of the deep crust can be inferred
by observations of the quiescent luminosity (\citealt{Rutledge1999}).
Hence this source provides a test of the strength and location of
heat sources in the crust. The question is whether the crust can reach
temperatures sufficiently hot to initiate thermally unstable \carbon\ fusion
in the short time available during the outburst. To check this, we
constructed a time-dependent model based on recent calculations of
the heating from electron capture reactions in the crust (\citealt{Gupta2007}).
We set the mass number to $A=40$, which yields a large heat release
($0.40\nsp\MeV$ per accreted nucleon) in the outer crust. For the
inner crust, we use the model of \citet{Haensel1990}. We integrated
the thermal diffusion equation using a standard method-of-lines formalism.
The microphysics is identical to that in \citet{Gupta2007}. The total
heat released in the crust during outburst is $1.8\nsp\MeV\usp\amu^{-1}$.

\begin{figure}
\includegraphics[bb=0bp 0bp 350bp 275bp,clip,width=1\columnwidth]{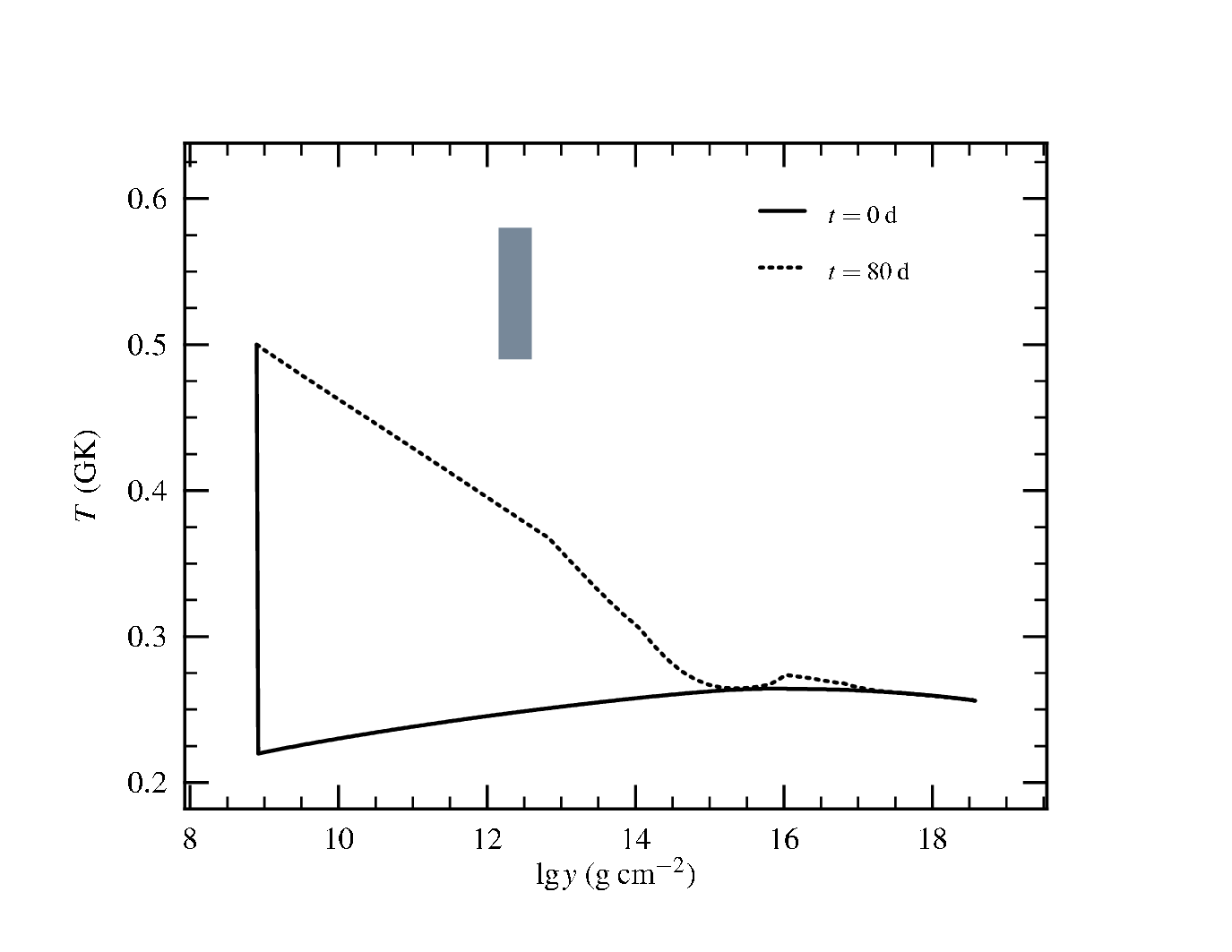}

\caption{\label{f.1608-crust}Temperature evolution in the crust of \bron~during
the outburst. At the beginning of the outburst (\emph{solid line})
we set the temperature at a column $y=10^{9}\,\mathrm{g\, cm^{-2}}$
to $5\cdot10^{8}\,\mathrm{K}$. After 80~days, the temperature in
the outer crust has risen (\emph{dotted line}), but is still well
below the temperature for which a one-zone stability analysis would
predict a superburst with a depth inferred from fits to the cooling
(\emph{shaded box}).}

\end{figure}

As an initial condition, we computed the temperature for steady accretion
at the time-averaged mass accretion rate $\langle\dot{M}\rangle=6.5\ee{15}\nsp\grampersecond$.
We then ran through a series of 60 outburst/quiescent cycles, to ensure
that the crust had reached a limit cycle. During outburst we set $\dot{M}=1.3\ee{17}\nsp\grampersecond$,
with a duration of 80\nsp\days; the quiescent interval was set to
4.16\nsp\yr. During quiescence we used a relation between the temperature
at a column $y=10^{9}\nsp\columnunit$ and the surface luminosity
appropriate for an accreted envelope (\citealt{Brown2002}). The quiescent
luminosity reached a minimum of $L_{q}=6.1\ee{33}\nsp\ergspersecond$,
in reasonable agreement with the value derived by \citet{Rutledge2000}.
Figure~\ref{f.1608-crust} shows the change in temperature, as a
function of column, during the outburst. At the start of the outburst
(\emph{solid line}), we set the temperature at a column $y=10^{9}\nsp\columnunit$
to $5\ee{8}\nsp\K$. This is a reasonable upper limit on the
temperature set by steady-state H/He burning. After 80\nsp\days,
the temperature in the outer crust has risen substantially (\emph{dotted
line}), but is still much cooler than needed to explain the fitted
ignition column and temperature (\emph{shaded box}).

We conclude that heating in the crust from electron captures, neutron
emissions, and pycnonuclear reactions, is insufficient to raise the
crust temperature during the outburst enough for \carbon\ to ignite
at the inferred column.

In the other superburst sources, which are either persistent accretors,
or transients with long duration outbursts (e.g. KS~1731-260), the
crust reaches a thermal steady state. Even in these sources, theoretical
models have difficulty explaining the inferred superburst ignition
columns of $\sim10^{12}\ \mathrm{{\rm g\ cm^{-2}}}$ (\citealt{Brown2004,Cooper2005,Cumming2006,Gupta2007}).
The properties of the neutron star crust and core that affect the
thermal profile such as neutrino emissivity and crust conductivity
must all be chosen to maximize the crust temperature in order to achieve
the observed ignition depths. In fact, \citet{Cumming2006} showed
that it was impossible to reproduce the observations if neutrino cooling
due to Cooper pair formation of superfluid neutrons in the crust (e.g.~\citealt{Yakovlev1999})
was included, because this efficient neutrino emission mechanism led
to an upper limit on the crust temperature at the ignition depth of
$\approx4\times10^{8}\ {\rm K}$, less than the temperature of $\approx6\times10^{8}\ {\rm K}$
required for carbon ignition. Recent work by \citet{Leinson2006}
has shown that the neutrino emission due to this mechanism has been
overestimated in previous calculations, and that this mechanism is in fact 
not the dominant source of neutrino emissivity.

As we have shown, the short duration of the transient outbursts in \bron\ means 
that the crust is heated to much lower temperatures than those corresponding to 
thermal equilibrium. Therefore the observation of a superburst from \bron\ is 
deeply puzzling and presents a significant challenge to current superburst models.

\subsection{Bursting behavior and mHz QPOs\label{sub:Bursts}}

The normal type-I X-ray bursting behavior looks similar to what is
seen in persistent bursting sources with strongly variable persistent
fluxes, most notably 4U~1820-303 (\citealt{cho1}), GX~3+1 (\citealt{Hartog2003}),
4U~1705-44 and KS~1731-260 (\citealt{cor02}). In those cases X-ray
bursts become far less frequent or even absent during high flux states.
Also in the case of \bron\ the bursting behavior is seen to differ
between the low and high flux state. \citet{Murakami1980} report
that in the high flux state bursts have a shorter duration and a higher
$\alpha$ value than the bursts in the low flux state, while the burst
rate is not significantly different. The bursts observed with the
RXTE PCA and the BeppoSAX WFCs exhibit this behavior as well. This
would be consistent with hydrogen and helium burning becoming predominantly
stable at higher mass accretion rates (\citealt{Paradijs1988}).

\citet{Revnivtsev2001} examined EXOSAT and RXTE PCA data of \bron\ and
discovered low frequency quasi-periodic oscillations (mHz QPOs). They
observed no type-I X-ray bursts when the persistent emission was higher
than when the QPOs were detected, which supported the hypothesis that
mHz QPOs occur at the transition of unstable to stable burning, such
that at higher accretion rates the bursting behavior ceases (see also
\citealt{Heger2005}). Interestingly, the PCA,
WFCs and IBIS/ISGRI have observed a total of four X-ray bursts when the persistent
flux was up to a factor of two higher than when mHz QPOs were seen (see
Fig.~\ref{fig:Number-of-observed}). Although this disagrees with the prediction of
the marginally stable burning model of \cite{Heger2005}, the hypothesis could
still be valid if, as discussed by \cite{Heger2005},
the conditions for nuclear burning vary substantially across the neutron star
surface, such that on one part of the surface unstable burning gives
rise to type-I bursts while on another part mHz QPOs occur. The scenario that
ignition conditions might vary across the surface of the star for rapidly
rotating neutron stars has been discussed by \cite{Cooper2007} (see also
\citealt{Spitkovsky2002}), although
the predicted range for the global accretion rate where both stable and unstable
burning take place at different latitudes is a factor of 3 higher than where mHz
QPOs are observed from \bron.

\section{Summary and conclusions}

For the first time a superburst has been observed from a transiently
accreting neutron star in a low-mass X-ray binary where the duration
of the accretion outbursts is short with respect to the expected superburst
recurrence time. \bron\ exhibited a superburst 55 days after the
onset of an accretion outburst. We analyzed the superburst as well
as the long term accretion and bursting behavior of \bron. We found
the properties of the superburst to be comparable to those of the
long superbursts observed from 4U~1254-690 and KS~1731-260. Fits of cooling 
models to the superburst light curve indicate that the superburst ignited 
when a column of $y=(1.5-4.1)10^{12}\,\mathrm{g\, cm^{-2}}$
had been accreted. From the average X-ray flux that the RXTE ASM observed
from \bron~over the past 11 years, we derive that this column was
accreted over a period of 26 to 72 years. 

The difficulties that superburst theory has in explaining the ignition
column depths derived for other superbursters, are underlined by the observation
of the superburst in \bron. 
The transient nature of the accretion poses challenges for superburst models. The average accretion
rate over the past 11 years was much lower than for the other superbursters, with the possible
exception of 4U~0614+091 (\citealt{kuu05}).
At the derived column depth, current superburst theory predicts that 
during the timespan of the outbursts the crust temperature does not rise to 
the value required for superburst ignition. Furthermore, the carbon that fuels 
the superburst is expected to be produced only during
the outbursts, while the stable nuclear burning of $^{12}$C as well as the 
frequent type-I X-ray bursts lower the carbon abundance. The detection of the
superburst implies that a significant amount of carbon survives the long periods outside the outbursts.
The balance between the creation and destruction processes is 
influenced by mixing and sedimentation in the neutron star envelope. The effect of these requires further
study before predictions about the produced amount of superburst fuel
can be made.

\begin{acknowledgements}
LK acknowledges support from the Netherlands Organization for Scientific
Research (NWO). Nuovo Telespazio and the BeppoSAX Science Data Center
are thanked for continued support. This research made use of the RXTE/ASM
database. AC is currently supported by an NSERC Discovery Grant, Le Fonds Qu\'eb\'ecois de la Recherche sur la Nature et les Technologies, and the Canadian Institute for Advanced Research, and is an Alfred P.~Sloan Research Fellow. EFB acknowledges support by Chandra Award Number TM7-8003X
issued by the Chandra X-ray Observatory Center, which is operated
by the Smithsonian Astrophysical Observatory for and on behalf of
the National Aeronautics Space Administration under contract NAS8-03060.
AC and EFB thank the Institute for Nuclear Theory at the University
of Washington for its hospitality and the Department of Energy for
partial support during the completion of this work.
\end{acknowledgements}
\bibliographystyle{aa}
\bibliography{sb1608}

\end{document}

%% file: aas_macros.tex
\def\jnl@style{\it}
\def\aaref@jnl#1{{\jnl@style#1}}
\def\aaref@jnl#1{{\jnl@style#1}}

\def\aj{\aaref@jnl{AJ}}                   % Astronomical Journal
\def\araa{\aaref@jnl{ARA\&A}}             % Annual Review of Astron and Astrophys
\def\apj{\aaref@jnl{ApJ}}                 % Astrophysical Journal
\def\apjl{\aaref@jnl{ApJ}}                % Astrophysical Journal, Letters
\def\apjs{\aaref@jnl{ApJS}}               % Astrophysical Journal, Supplement
\def\ao{\aaref@jnl{Appl.~Opt.}}           % Applied Optics
\def\apss{\aaref@jnl{Ap\&SS}}             % Astrophysics and Space Science
\def\aap{\aaref@jnl{A\&A}}                % Astronomy and Astrophysics
\def\aapr{\aaref@jnl{A\&A~Rev.}}          % Astronomy and Astrophysics Reviews
\def\aaps{\aaref@jnl{A\&AS}}              % Astronomy and Astrophysics, Supplement
\def\azh{\aaref@jnl{AZh}}                 % Astronomicheskii Zhurnal
\def\baas{\aaref@jnl{BAAS}}               % Bulletin of the AAS
\def\jrasc{\aaref@jnl{JRASC}}             % Journal of the RAS of Canada
\def\memras{\aaref@jnl{MmRAS}}            % Memoirs of the RAS
\def\mnras{\aaref@jnl{MNRAS}}             % Monthly Notices of the RAS
\def\pra{\aaref@jnl{Phys.~Rev.~A}}        % Physical Review A: General Physics
\def\prb{\aaref@jnl{Phys.~Rev.~B}}        % Physical Review B: Solid State
\def\prc{\aaref@jnl{Phys.~Rev.~C}}        % Physical Review C
\def\prd{\aaref@jnl{Phys.~Rev.~D}}        % Physical Review D
\def\pre{\aaref@jnl{Phys.~Rev.~E}}        % Physical Review E
\def\prl{\aaref@jnl{Phys.~Rev.~Lett.}}    % Physical Review Letters
\def\pasp{\aaref@jnl{PASP}}               % Publications of the ASP
\def\pasj{\aaref@jnl{PASJ}}               % Publications of the ASJ
\def\qjras{\aaref@jnl{QJRAS}}             % Quarterly Journal of the RAS
\def\skytel{\aaref@jnl{S\&T}}             % Sky and Telescope
\def\solphys{\aaref@jnl{Sol.~Phys.}}      % Solar Physics
\def\sovast{\aaref@jnl{Soviet~Ast.}}      % Soviet Astronomy
\def\ssr{\aaref@jnl{Space~Sci.~Rev.}}     % Space Science Reviews
\def\zap{\aaref@jnl{ZAp}}                 % Zeitschrift fuer Astrophysik
\def\nat{\aaref@jnl{Nature}}              % Nature
\def\iaucirc{\aaref@jnl{IAU~Circ.}}       % IAU Cirulars
\def\aplett{\aaref@jnl{Astrophys.~Lett.}} % Astrophysics Letters
\def\apspr{\aaref@jnl{Astrophys.~Space~Phys.~Res.}}
\def\bain{\aaref@jnl{Bull.~Astron.~Inst.~Netherlands}} 
\def\fcp{\aaref@jnl{Fund.~Cosmic~Phys.}}  % Fundamental Cosmic Physics
\def\gca{\aaref@jnl{Geochim.~Cosmochim.~Acta}}   % Geochimica Cosmochimica Acta
\def\grl{\aaref@jnl{Geophys.~Res.~Lett.}} % Geophysics Research Letters
\def\jcp{\aaref@jnl{J.~Chem.~Phys.}}      % Journal of Chemical Physics
\def\jgr{\aaref@jnl{J.~Geophys.~Res.}}    % Journal of Geophysics Research
\def\jqsrt{\aaref@jnl{J.~Quant.~Spec.~Radiat.~Transf.}}
\def\memsai{\aaref@jnl{Mem.~Soc.~Astron.~Italiana}}
\def\nphysa{\aaref@jnl{Nucl.~Phys.~A}}   % Nuclear Physics A
\def\physrep{\aaref@jnl{Phys.~Rep.}}   % Physics Reports
\def\physscr{\aaref@jnl{Phys.~Scr}}   % Physica Scripta
\def\planss{\aaref@jnl{Planet.~Space~Sci.}}   % Planetary Space Science
\def\procspie{\aaref@jnl{Proc.~SPIE}}   % Proceedings of the SPIE